\documentclass[final]{siamltex}
\usepackage{geometry,enumerate}                
\usepackage{graphicx}
\usepackage{amssymb,amsmath}
\usepackage{epstopdf}

\newcommand{\unit}[1]{\,\scriptstyle{\ensuremath{\, \mathrm{#1}}}}
\usepackage{showkeys}
\title{The cellular dynamics of bone remodeling: \\ a mathematical model}
\author{Marc D. Ryser\thanks{Department of Mathematics and Statistics, McGill University, Montreal, Canada ({\tt ryser@math.mcgill.ca}).}
        \and Svetlana V. Komarova \thanks{Faculty of Dentistry, McGill University, Montreal, Canada}
        \and Nilima Nigam \thanks{Department of Mathematics, Simon Fraser University, Burnaby, Canada}
        }

\begin{document}
\maketitle

\begin{abstract}
The mechanical properties of vertebrate bone are largely determined by a process which involves the complex interplay of three different cell types. This process is called {\it bone remodeling}, and occurs asynchronously at multiple sites in the mature skeleton. The cells involved are bone resorbing osteoclasts, bone matrix producing osteoblasts and mechanosensing osteocytes. These cells communicate with each other by means of autocrine and paracrine signaling factors and operate in complex entities, the so-called bone multicellular units (BMU). To investigate the BMU dynamics in silico, we develop a novel mathematical model resulting in a system of nonlinear partial differential equations with time delays. The model describes the osteoblast and osteoclast populations together with the dynamics of the key messenger molecule RANKL and its decoy receptor OPG. Scaling theory is used to address parameter sensitivity and predict the emergence of pathological remodeling regimes. The model is studied numerically in one and two space dimensions using finite difference schemes in space and explicit delay equation solvers in time. The computational results are in agreement with in vivo observations and provide new insights into the role of the RANKL/OPG pathway in the spatial regulation of bone remodeling.

\end{abstract}



\pagestyle{myheadings}
\thispagestyle{plain}
\markboth{Ryser, Komarova, Nigam}{Mathematical Model of Bone Remodeling}

\section{Introduction}

The vertebrate skeleton plays a crucial role in providing mechanical support as well as a ready source of calcium and other important minerals. Physical loading of the skeleton causes stresses which can lead to local micro-damage in the bone tissue. Similarly, if the calcium level in the blood drops below a certain threshold, systemic regulators such as hormones transmit the order to release calcium through removal ({\it resorption}) of bone tissue. In both cases, the resorbed spaces have to be filled with sound tissue in order to restore the structural integrity. This joint process of bone destruction and re-growth is referred to as \emph{bone remodeling}, and is realized by complex multicellular entities, the so-called \emph{bone multicellular units} (BMU). Each BMU consists of several interacting cell types and a whole variety of biochemical signaling factors.  The importance of remodeling becomes apparant when considering the implications of its malfunctioning. Deficient or even absent remodeling of micro-damage can lead to macroscopic bone fractures, and pathologies in the BMU functioning are largely responsible for diseases such as osteoporosis and rheumatoid arthritis \cite{Kong1999Activated}.\\
\indent The various physiological and pathological aspects of BMUs have been studied by both experimentalists and clinicians for well over 40 years \cite{Parfitt1994Osteonal}. However, due to a general lack of conclusive \emph{in vivo} experiments \--- so far mainly consisting of histological sections of dead bone tissue \--- several phenomena remain poorly understood. The difficulty and costs for \emph{in vivo} experiments suggest that there is great potential for mathematical modeling in this field. So far, several research groups have modeled the local strain fields in bones \cite{Smit2000BMUcoupling, Yosibash2007Reliable}  as well as the temporal sequence of local bone destruction and re-growth at the cellular level \cite{Komarova2005Mathematical, Komarova2003Mathematical, Lemaire2004Modeling}. In essence, the latter  models successfully capture the local bone cell dynamics in physiological settings and are even able to describe certain pathologies. However, the functioning of a remodeling unit strongly depends on its spatial organization and therefore, a purely temporal model cannot  provide a complete description of the BMU. To address this, we develop here a novel spatio-temporal model of a single remodeling unit, describing the dynamics of both the involved bone cell populations as well as the relevant signaling pathways. The model consists of five nonlinear partial differential equations (PDE) and is based on a continuum assumption for the cell populations. \\
\indent In section \ref{bio}, we first give an outline of the relevant biology, thereby focusing on the three types of bone cells (osteoclasts, osteoblasts, osteocytes) and the most important biochemical factors (the  RANK/RANKL/OPG pathway). Once these concepts are established, we begin the model development in section \ref{mathmod} by introducing a previous temporal model by Komarova et al. \cite{Komarova2005Mathematical, Komarova2003Mathematical}.  Given the complexity of the underlying biological system \--- involving endocrine signaling,  cell motion, fluid diffusion etc. \--- some simplifying assumptions are necessary in order to develop a compact and closed spatio-temporal model. The model is developed in an abstract setting independent of the spatial dimension, but can be applied to one (1D), two (2D) or three (3D) dimensions.  In section \ref{1Dmodel} we present the 1D case, use scaling theory to gain insight into parameter sensitivity, and present experiments focusing on the different pathological regimes. The biologically more relevant 2D case is then discussed in section \ref{2Dmodel} Êand a selection of two physiological remodeling experiments is presented. The results of the 2D experiments provide a model validation as well as new insights into the role of the RANK/RANKL/OPG pathway in the spatial regulation of bone remodeling.

\section{The biology of bone remodeling}\label{bio}

Bone remodeling refers to the combination of bone destruction and subsequent re-growth. It is a coordinated process of three different cell types that interact by means of several biochemical factors. Furthermore, mechanical strains play an important role in the stimulation and steering of remodeling units.  The following outline is focused on the model-relevant mechanisms and we refer to \cite{Parfitt1994Osteonal, Robling2006Biomechanical} for detailed reviews.

\subsection{The bone cells}
Three different cell types are involved in remodeling.
\begin{list}{\labelitemi}{\leftmargin=1em}

\item {\bf Osteoclasts}   \cite{Quinn2001Transforming, Boyle2003Osteoclast} are cells which resorb mineralized bone tissue while moving along the bone surface. They are formed by cell differentiation from stem cells in the bone marrow and have a life span of roughly 10 days. A key stimulator for osteoclast differentiation and activation is a molecule called RANKL (the receptor activator of nuclear factor $\kappa$B). 

\item {\bf Osteoblasts}  \cite{Harada2003Control} are cells which fill the previously resorbed trench with osteoid, the organic part of the bone tissue. Later on, osteoid mineralizes and the remodeling process is complete. Osteoblasts differentiate from stem cells in the bone marrow, they do not move along the bone surface, and they express the messenger molecule RANKL and its decoy receptor OPG (osteoprotegerin). After approximately two weeks, osteoblasts either die or differentiate into osteocytes and get buried alive in the new bone tissue.

\item {\bf Osteocytes} \cite{FranzOdendaal2006Buried, Bonewald2008Osteocytes} differentiate from active osteoblasts and are connected with each other to form a large network of active cells within the bone tissue. This network is believed to propagate information, to localize damage sites and micro-strains, and to play an important role in the process of mechanotransduction. 
\end{list}\vspace{.05cm}

\begin{figure}[!h]
\begin{center}
  \includegraphics[width=.4\textwidth]{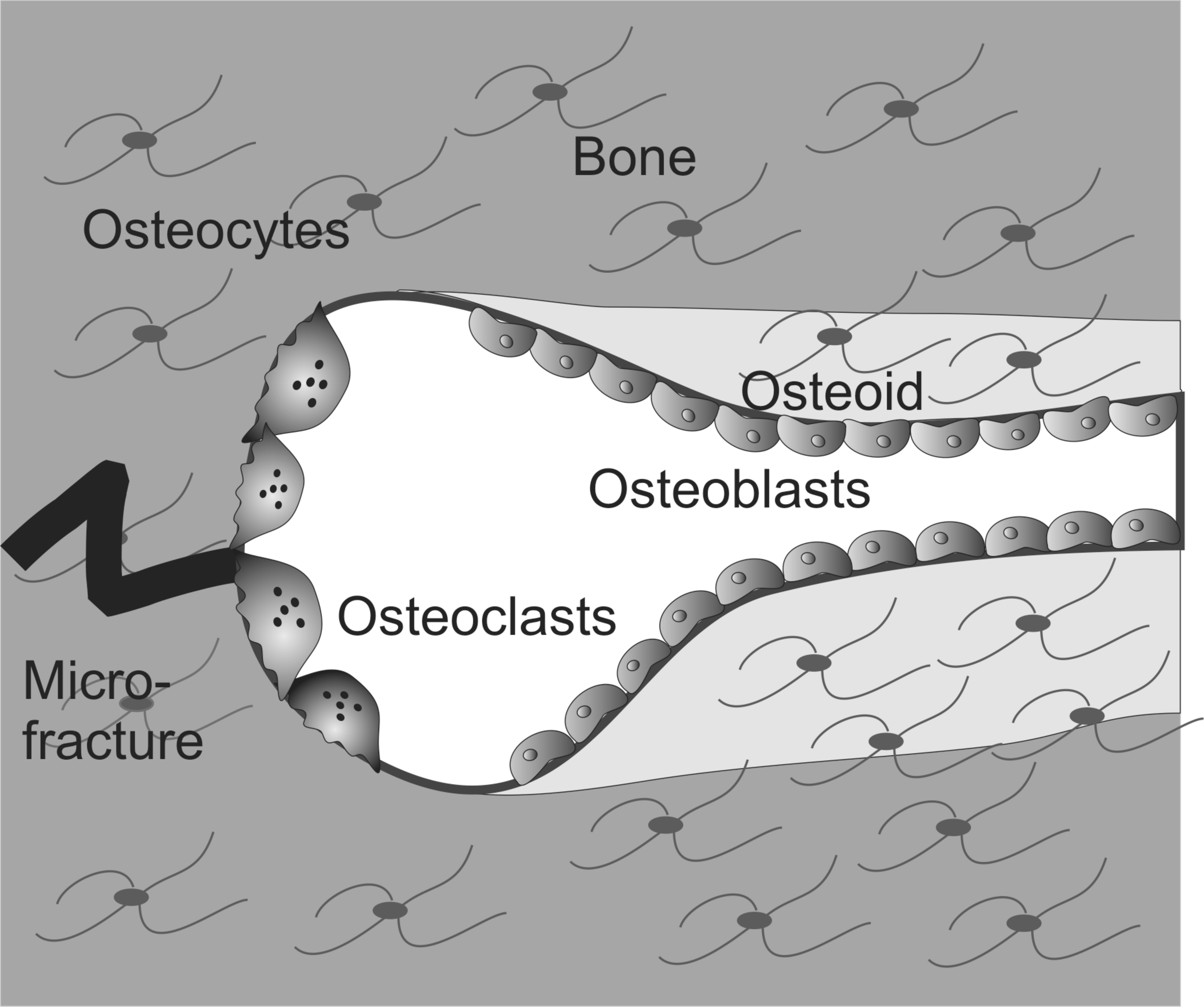}
  \label{fig:PP0}
	\caption{\footnotesize A schematic, not-to-scale representation of a BMU moving along a micro-fracture on a piece of trabecular bone. Osteoclasts resorb the bone in form of a cutting cone and osteoblasts subsequently fill the resorbed space with new bone matrix. Bone cells interact by means of cytokines and growth factors and osteoblasts differentiate into osteocytes.}
\end{center}
\end{figure}

\noindent The three cell types communicate by means of autocrine signaling (communication among cells of the
same type) and paracrine signaling (communication among cells of different types). Generally, the bone cells and their messengers operate locally in well-confined remodeling units, the BMUs. These units operate for up to 12 months in a row, thereby by far exceeding the individual cell's life spans. The progression of a BMU across the bone can be summarized as follows:
\begin{enumerate}[Step 1)]
\item Initially, $10-20$ osteoclasts are recruited to the initiation site and resorb the old bone tissue. Once the tissue is removed, the osteoclasts move on and keep on resorbing while traveling at a speed of $20-40 \mu m$ per day \cite{Jones2006Regulation, Parfitt1994Osteonal}. During the whole remodeling process, they stay together in a spatially well-confined aggregation (\emph{cutting cone}). Dead cells are continually replaced by new ones so that the population size remains approximately constant. 
\item Once the osteoclasts have resorbed the bone tissue, they recruit  1000-2000 osteoblasts that fill the previously resorbed trench with new bone matrix (\emph{closing cone}). Osteoblasts are much less efficient than osteoclasts and the bone formation takes roughly 10 times longer than the resorption. 
\item Finally, the new bone matrix mineralizes and osteoblasts either die or differentiate into osteocytes.
\end{enumerate}
There are two kinds of bone tissues. \emph{Cortical} tissue is dense and compact and forms the outer surface of bones. \emph{Trabecular} tissue fills the inner cavity with a honeycomb-like structure, consisting of irregularly shaped spicules (\emph{trabeculae}) endowed in bone marrow. Remodeling takes place in both cortical and trabecular bone and the difference in the respective BMU progressions is geometrical rather than biological in nature: whereas the BMU has to dig a complete tunnel to penetrate the compact cortical tissue, it can move along the surface of the trabeculae, thereby only digging a half-trench. Figure \ref{fig:PP0} illustrates the temporal sequence of the remodeling steps on a trabecula. 

\subsection{The biochemical factors involved in remodeling}

\begin{figure}[htbp] 
   \centering
   \includegraphics[width=0.6\textwidth]{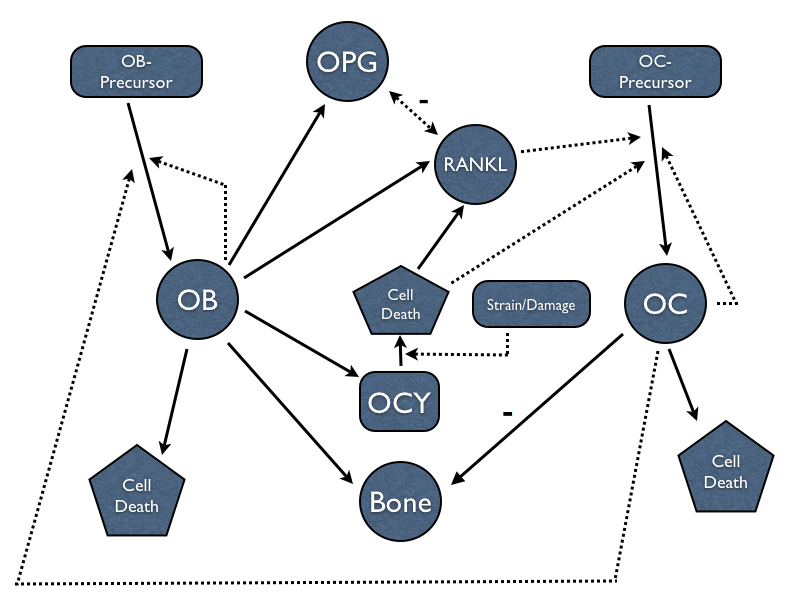} 
   \caption{Cells and biochemical factors known to play a role in the remodeling process of bone. The cells are osteoclasts (OC), osteoblasts (OB), osteocytes (OCY), and their respective precursor cells. Solid lines stand for positively balanced processes (cell differentiation and production of chemicals/tissue) and dotted lines for positively balanced regulations (autocrine/paracrine stimulation). The (-) next to an arrow indicates a negatively balanced process or regulation.}  \label{fig:PP1}
\end{figure}

The coordination of osteoclasts, osteoblasts and osteocytes within a BMU is realized through a sophisticated communication system, which consists of various autocrine and paracrine signaling pathways involving numerous coupled effectors. However, the multiple actions attributed to some of these effectors make it hard to identify the actual key players and to predict the cumulative dynamics of the coupling. Figure \ref{fig:PP1} summarizes the major control pathways in the remodeling process and identifies the respective messenger molecules. Among the multiple messengers involved, RANKL and OPG have been shown to play critical roles in both physiological bone remodeling and in the development of diseases \cite{Kong1999Activated, Bucay1998Osteoprotegerin, Kearns2008Receptor}.  RANKL is a cytokine produced in either membrane-bound or soluble form by cells of the osteoblast lineage, prominently by osteocytes and osteoblasts. Several studies have shown that RANKL is up-regulated in situations associated with increased bone remodeling, such as PTH treatment \cite{Huang2004PTH}, mechanical stimulation \cite{Kanzaki2002Periodental}, as well as fractures \cite{Ito2005Remodeling}. 
RANKL binds to RANK receptors on the surface of osteoclastic cells and has a stimulatory impact on the differentiation of osteoclast precursors and the subsequent activation of mature osteoclasts into active, resorbing cells. On the other hand, the molecule OPG, produced by mature osteoblasts \cite{Gori2000Expression}, acts as a decoy receptor of RANKL, i.e. it inhibits RANKL by forming RANKL-OPG complexes. Since the presence of OPG means less RANKL-RANK binding and hence less osteoclast stimulation, a high RANKL/OPG ratio favors bone resorption whereas a low ratio down-regulates osteoclastic activity. The RANK/RANKL/OPG pathway is also known to be employed by systemic regulators such as parathyroid hormone (PTH) and vitamin D to regulate the resorption activity. Note finally that the spatial separation of the different RANKL and OPG sources indicates that in addition to the local ratio of the chemicals, their spatial distribution plays an important role, too.

\subsection{The mechanical effects: microscopic strains and fractures}\label{mecheff}
There are two different remodeling modes, \emph{targeted}Ê and \emph{random} remodeling. Whereas the former mode aims at damage removal by means of local micro-fracture reparation, the latter serves the purpose of damage prevention: old \--- but not necessarily damaged \--- tissue is continually renewed across the skeleton to prevent fatigue damage. Both remodeling types rely on \emph{steering} mechanisms that ensure that BMUs are guided towards damage sites and move in a way that minimizes structural instability due to ongoing bone erosion. The concept of \emph{targeted steering} is based on established evidence that the presence of micro-fractures leads to creation of new BMUs and attraction of already existing BMUs \cite{Burr1997Muscle, Burr1993Calcified}. On the other hand, it has been suggested that strain-derived canalicular fluid flow is responsible for osteoclast activity and motility in the cutting cone of the BMU \cite{Burger2003Strainderived}, leading to \emph{strain-derived steering}. In particular, this steering mechanism ensures that BMUs move along the principal strain axis of the bone and hence optimize its robustness at anytime of the remodeling process. Both steering mechanisms rely on mechanical features that need to be translated into cell signals to attract BMUs. Recent investigations show that there is a unifying mechanism of mechanotransduction for both damage and strain, mediated by osteocytes. In fact, both mechanically damaged osteocytes and osteocytes exposed to fluid shear stress have been shown to express RANKL \cite{Bakker2004Shear, Kurata2006Bone, You2008Osteocytes}. Since RANKL is a potent osteoclast stimulator, this allows mechanically stimulated osteocytes to attract BMUs and hence guide them towards damage sites and along the principal stress directions.

\section{The Mathematical Model}\label{mathmod}
In this section, we develop a mathematical model describing the spatio-temporal evolution of a single BMU at cellular level. The overall \emph{goals of this model} are the following:
\begin{itemize}
\item To describe the distinctive spatial and temporal features of the cutting cone and the BMU movement.
\item To link the key biochemical factors RANKL and OPG with the known population dynamics of bone cells.
\item To test the model on experimental findings and suggest new experimental studies. 
\end{itemize}
\vspace{.05cm}
Since we develop a model that can be considered in one, two and three space dimensions, we do not specify its dimension explicitly and denote it simply by $n$, where $n=1,2,3$. The 1D and 2D versions of the model presented in this article are particularly suited for the description of \emph{trabecular remodeling}, and the restrictions of their applicability to cortical bone will be discussed in section \ref{concl}. The major modeling assumptions can be summarized as follows:

\begin{itemize}
\item We focus on trabecular remodeling, more precisely on the dynamics of a BMU moving across a single trabecula.
\item The trabecula is locally flat enough so that we can neglect  curvature.
\item We make a continuum assumption for the cell population densities, i.e., we shall be modeling cell densities rather than individual cells.  
\item The BMU evolves along the surface of the trabecula and the depth of the resorbed trench ($\sim 10\mu m$) is small in comparison to its width ($\sim 500\mu m$).
\item Of the several cell types involved in remodeling \--- osteoblasts, osteoclasts, osteocytes and their respective precursors \--- we only consider osteoblasts and osteoclasts as state-variables.
\item  The trabecula is endowed in bone marrow which can be considered as a reservoir of precursor cells.
\item Among the multitude of biochemical factors, only RANKL and OPG  are modeled explicitly, the rest of the factors \--- such as TGF-$\beta$, IGF, M-CSF and nitric oxide \--- are captured in nonlinear interactions. 
\item The canopy of bone lining cells separating the BMU from the bone marrow \cite{Hauge2001Cancellous} ensures that the loss of chemicals by vertical diffusion is negligible.
\item We model the elimination of OPG and RANKL through their mutual interaction only and do not include their natural decay rates.
\item The mechanical factors responsible for the BMU steering \--- microscopic strains and damages \--- are modeled implicitly in form of an appropriate RANKL distribution in the initial field. For the sake of simplicity, we will from now on refer to these distributions as micro-fractures, even though they might be caused by local micro-strains, see section \ref{mecheff}.
\end{itemize}
\noindent Due to the complexity of the model, we proceed in three steps, starting off with a brief review of the temporal model introduced by \cite{Komarova2005Mathematical, Komarova2003Mathematical}. In a second step, we introduce the spatial extension of the model as well as the RANKL and OPG fields. In a third step, we complete the model by adding appropriate initial and boundary conditions.

\subsection{Prior work: temporal model}\label{priwo}
The model suggested by Komarova et al. \cite{Komarova2005Mathematical, Komarova2003Mathematical} is a temporal model describing the population dynamics of bone cells at a single point within the BMU. Denoting the number of osteoclasts and osteoblasts by $u_1$ and $u_2$ respectively, the cell dynamics are modeled as 
 \begin{equation}\label{temp1}
  \left\{ \begin{array}{lll}
 \partial_t u_1&=&\alpha_1 u_{1}^{g_{11}} u_{2}^{g_{21}}-\beta_1 u_1,  \\
 \partial_t u_2&=&\alpha_2 u_{1}^{g_{12}} u_{2}^{g_{22}}-\beta_2 u_2,
\end{array} \right.
 \end{equation}
 where $\alpha_i$ and $\beta_i$ are activities of cell production and death and all have units $[day^{-1}]$. The four dimensionless parameters $g_{ij}$ represent the effectiveness of the autocrine and paracrine interactions between the constituent cells. Let us now briefly discuss the various signaling factors $g_{ij}$, thereby making some restrictions appropriate to the spatio-temporal model we are finally aiming for. The factor $g_{11}$ represents the effectiveness of the autocrine interactions between osteoclasts and has been shown to control the overall remodeling dynamics \cite{Komarova2003Mathematical}. Osteoclast-derived paracrine regulation of osteoblasts ($g_{12}$) is the crucial link in the BMU coupling and its inhibition leads to negatively balanced remodeling \cite{Komarova2005Mathematical}. Regarding the autocrine stimulation of osteoblasts ($g_{22}$), it is known that the latter express auto-stimulatory factors such as insulin-like growth factors IGF \cite{Canalis1996Calcified}. However, these factors do not influence the dynamical behavior of the BMU \cite{Komarova2003Mathematical} and we assume here that they are negligible in comparison to the impact of $g_{12}$, i.e  we set $g_{22}=0$. Finally, osteoblast-derived paracrine regulation of osteoclasts is dominated by the RANK/RANKL/OPG pathway \cite{Yasuda1998Osteoclast, Khosla2001Minireview} and therefore the factor $g_{21}$ plays an important role in the temporal model. However, since we will eventually develop a model that includes the RANKL and OPG fields explicitly as state-variables, we can set $g_{21}=0$. After these simplifications, the system (\ref{temp1}) reduces to 
 \begin{equation}\label{temp1alpha}
  \left\{ \begin{array}{lll}
 \partial_t u_1&=&\alpha_1 u_{1}^{g_{11}} -\beta_1 u_1,  \\
 \partial_t u_2&=&\alpha_2 u_{1}^{g_{12}}-\beta_2 u_2.
\end{array} \right.
 \end{equation}
For $g_{11}<1$, the unique non-trivial fixed point $(u_{1,ss},u_{2,ss})>(0,0)$ of equation (\ref{temp1alpha}) is a stable node. It is assumed that cells below the steady-state values $u_{i,ss}$ are precursor cells which are  less differentiated. In other words, they are not actively involved in the resorption and production of bone matrix, but participate in autocrine and paracrine signaling. Increases in $u_i$ above $u_{i,ss}$ are regarded as proliferation and differentiation of precursors into mature osteoclasts and osteoblasts that participate actively in the remodeling process. In this sense, the initiation of remodeling can be induced manually by increasing the number of osteoclasts above the equilibrium value, i.e. by choosing initial conditions $u_1(t_0)>u_{1,ss}$. Note that $u_2(t_0)=u_{2,ss}$ is sufficient because it is assured that osteoblasts are recruited by active osteoclasts. 
For all the subsequent numerical experiments we will choose the parameter $g_{11}<1$ such that $(u_{1,ss}, u_{2,ss})$ corresponds to a stable steady-state solution of (\ref{temp1alpha}). Together with the initiation procedure explained above, this implies that $(u_{1}, u_{2})\geq (u_{1,ss}, u_{2,ss})$ for all $t\geq t_0$ and hence we can ensure that the populations of active cells, denoted hereinafter by $y_i\equiv u_i-u_{i,ss}$,  remain non-negative. Using the decomposition $u_i\equiv u_{i,ss}+y_i$, we can see that the system (\ref{temp1alpha}) actually describes the evolution of the active cell populations coupled to the constant precursor populations 
\begin{equation}\label{temp2alpha}
  \left\{ \begin{array}{lll}
 \partial_t y_1&=&\alpha_1 (u_{1,ss}+y_1)^{g_{11}} -\beta_1 (u_{1,ss}+y_1), \\
 \partial_t y_2&=&\alpha_2 (u_{1,ss}+y_1)^{g_{12}}-\beta_2 (u_{2,ss}+y_2).
\end{array} \right.
 \end{equation}
Even though our main interest is the evolution of the active cells in (\ref{temp2alpha}), we will henceforth use the equivalent version (\ref{temp1alpha}) for its more compact notation. The active cell populations are then easily recovered by subtracting the corresponding precursor populations $u_{i,ss}$ from the solutions $u_i$ of (\ref{temp1alpha}).

\subsection{The spatial extension}\label{spatex}
We use now the temporal model (\ref{temp1}) as the basis for the spatial extension. The model developments in this section are independent of the spatial dimension and we avoid a specific choice by denoting all differential operators by their multidimensional symbols such as $\nabla$ and $\Delta$. Later on we will discuss the 1D case in section \ref{1Dmodel} and the 2D case in section \ref{2Dmodel}. The units of the parameters introduced below can all be found in the Appendices \ref{paramval1D} and \ref{paramval2D}. 

First, we switch to space-dependent state variables $u_i(t) \mapsto u_i(\boldsymbol{x},t)$, where $\boldsymbol{x}\in \Omega \subset \mathbb{R}^n$  and the domain $\Omega$ is chosen large enough to avoid interactions with the boundaries ($n=1,2,3$ is the spatial dimension). Note in particular that the $u_i$ have now the units of a density $[ mm^{-n}]$. At the same time we introduce the RANKL and OPG fields as new state variables. They are denoted by $\phi_R(\boldsymbol{x},t)$ and $\phi_O(\boldsymbol{x},t)$ and have the units of a concentration $[mol \ mm^{-n}]$. To build up the final model we proceed now in two steps. First, we assume that the RANKL and OPG fields are known and analyze their impact on osteoclasts and osteoblasts. In a second step we introduce then the equations governing the spatio-temporal evolution of the RANKL and OPG fields themselves.  Finally, we would like to emphasize that throughout the spatial extension the quantities $u_{i,ss}$ refer to the steady-state densities of the temporal equation (\ref{temp1alpha}) and \textit{not} Êto the steady-state solutions of the spatial equations.

\subsubsection{The impact of RANKL and OPG on osteoclasts and osteoblasts}
RANKL is known to have an important impact on osteoclasts: it promotes their differentiation and activation and contributes together with other signaling molecules to the navigation (\emph{chemotaxis}) of active cells \cite{Boyle2003Osteoclast, Jones2006Regulation}. On the other hand, the only impact of OPG on osteoclasts is indirect by means of RANKL inhibition. Accordingly, the osteoclast equation in  (\ref{temp1}) has to be augmented by two contributions only:
\begin{equation}\label{OCspat}
 \partial_t u_1=\alpha_1 u_{1}^{g_{11}} -\beta_1 u_1  -\underbrace{\zeta \,\nabla  \cdot (y_1 \nabla  \phi_R)}_{C1}+\underbrace{k_1\ \frac{\phi_R}{\lambda+\phi_R} \  \theta(y_1)\ u_1}_{C2}.
 \end{equation}
The term $C1$ describes the motion of active osteoclasts along the gradient of the RANKL field and $\zeta$ indicates the effectiveness of migration.
The second term $C2$ represents the stimulating action of RANKL on osteoclasts via RANK-RANKL binding ($k_1$ is the corresponding reaction rate). This comprises both the differentiation of precursor cells into active osteoclasts as well as the steadily occurring renewal of nuclei in already resorbing cells \cite{Parfitt1994Osteonal}. We assume that the RANK receptors have a saturation threshold, hence the sigmoid function with $\lambda$ as the concentration of half-saturation. The Heaviside function $\theta(y_1)$, defined as $\{\theta(x)=0 \textrm{ if $x\leq0$, }\ \theta(x)=1 \textrm{ if $x>0$ }\}$, ensures that stimulation takes place only in presence of active osteoclasts ($y_1$), i.e. only osteoclasts ($u_1$) in the cutting cone area are stimulated by RANKL. It is easy to verify that if $u_1(t_0)\geq u_{1,ss}$, then $u_1\geq u_{1,ss}$ for all $t\geq t_0$, i.e. the population of active osteoclasts stays non-negative. Therefore, the same comments as in section \ref{priwo} apply and equation (\ref{OCspat}) can, similarly to equations (\ref{temp1alpha}) and (\ref{temp2alpha}), be rewritten as an evolution equation for $y_1$. \\ \indent Regarding osteoblasts, we assume that they are recruited by osteoclasts and do not move by themselves. Since RANKL and OPG have no significant impact on their dynamics, the $u_2$ equation in (\ref{temp1}) remains unaltered.

 \subsubsection{Dynamics of RANKL and OPG fields}
The evolution of the RANKL  concentration $\phi_R$ is governed by production, diffusion and reaction. More precisely, RANKL is expressed by active osteoblasts, it spreads across the trabecula through diffusion and it binds to OPG as well as RANK receptors on osteoclasts. In mathematical terms, the rate of change in time reads
\begin{eqnarray}\label{phiR1}
\partial_t \phi_R &=& \underbrace{a_R\ y_{2,t_R}}_{C3}+\underbrace{\kappa_R \Delta(\phi_{R}^{\epsilon_R})}_{C4} -\underbrace{k_2\ \frac{\phi_R}{\lambda+\phi_R} \  \theta(y_1)\ u_1}_{C5}-\underbrace{k_3\phi_R\phi_O}_{C6}.
\end{eqnarray}
The RANKL source by active osteoblasts $C3$ is justified as follows: after the differentiation of precursors into mature osteoblasts, it takes a certain time $t_R$ until the cells start to produce RANKL \cite{Gori2000Expression, Thomas2001Changing, Baldock2006Vitamin}. The number of active osteoblasts at time $t$ that are of age $t_R>t$ or older is $e^{-\beta_2 t_R } y_2(\boldsymbol{x},t-t_R)$ and after absorbing the constant prefactor into the proportionality constant $a_R$ we obtain $C3$, where $y_{2,t_R}\equiv y_2(\boldsymbol{x},t-t_R)$. The second contribution $C4$ takes care of the porous diffusion which can vary between very low for membrane-bound RANKL and high for soluble RANKL.  $\kappa_R$ is the diffusion constant and the dimensionless exponent $\epsilon_R\geq1$ reflects the porosity of the medium surrounding the BMU. Note that if $\epsilon_R > 1$, then an initially compactly supported RANKL field will stay compactly supported over time; this is not the case for the regular diffusion equation which is known to have infinite propagation speed. Since the BMU environment is very irregular and since the spreading cytokines are in steady interaction with the various constituents of the bone matrix as well as adjacent bone cells, the porous version with $\epsilon_R>1$ seems to provide a more plausible model for the RANKL field than the regular version with $\epsilon_R=1$. For a more detailed discussion of porous medium equations we refer to \cite{Evans1998Partial, Ebmeyer2005Regularity}. The contribution $C5$ is due to the receptor-ligand binding of RANK and RANKL and is almost identical to $C2$ in equation (\ref{OCspat}), except for the different rate constant $k_2$. Note that $k_2$ contains information about several factors such as receptor density on osteoclasts and reversibility of the RANK-RANKL binding. Finally, the reaction term $C6$ models the RANKL-OPG binding with rate constant $k_3$.  \\
\indent Similarly to $\phi_R$, the rate of change in the OPG field $\phi_O$ is also governed by the contributions of source, diffusion and reaction:
 \begin{eqnarray}\label{OPG1}
 \partial_t \phi_O&=&\underbrace{a_O\ y_{2,t_O}}_{C7}+\underbrace{\kappa_O \Delta(\phi_O^{\epsilon_O})}_{C8} -\underbrace{k_3\phi_R\phi_O}_{C9}.
 \end{eqnarray}
Similarly to $C3$ in equation (\ref{phiR1}), OPG is produced by mature osteoblasts with a time delay $t_O$ such that $t_O > t_R$  \cite{Gori2000Expression, Thomas2001Changing, Baldock2006Vitamin}. The contribution $C8$ for porous diffusion ($\epsilon_O\geq 1$) is analog to $C4$ and the OPG-RANKL binding $C9$ is identical to $C6$. Note that the diffusion parameters of RANKL ($\kappa_R$, $\epsilon_R$) and OPG ($\kappa_O$, $\epsilon_O$) are not necessarily equal. In a physiological setting, RANKL is mainly membrane bound whereas OPG is soluble.

\subsection{The complete model}\label{finalmod}

Together with the evolution of the bone density $z(\boldsymbol{x},t)$ \--- diminished by active osteoclasts and augmented by active osteoblasts \--- equations (\ref{temp1}), (\ref{OCspat}), (\ref{phiR1}) and (\ref{OPG1}) yield the following nonlinear, time-delayed partial differential equation
\begin{equation} \left\{ 
              \begin{array}{lll}\label{PDE}
               \partial_t u_1&=&\alpha_1 u_{1}^{g_{11}} -\beta_1 u_1 - \zeta \nabla \cdot (y_1 \nabla \phi_R)+k_1\ \ \frac{ \phi_R}{\lambda+\phi_R} \  \theta(y_1)\ u_1 \\
\partial_t u_2&=&\alpha_2 u_{1}^{g_{12}} -\beta_2 u_2 \\
\partial_t \phi_R &=& a_R\ y_{2,t_R} +\kappa_R \Delta (\phi_{R}^{\epsilon_R})-k_2\ \frac{ \phi_R}{\lambda+\phi_R} \  \theta(y_1)\ u_1-k_3\phi_R\phi_O\\
\partial_t \phi_O&=&a_O\ y_{2,t_O}  +\kappa_O \Delta(\phi_O^{\epsilon_O})-k_3\phi_R\phi_O\\
\partial_t z&=&-f_1\ y_1+f_2\ y_2.
                                 \end{array}
       \right. 
\end{equation}
Recall that $y_i\equiv u_i-u_{i,ss}$ are the active cells and $y_{2,t_\omega}\equiv y_2(\boldsymbol{x},t-t_\omega)$. The mechanisms behind BMU initiation are still not fully understood and we do not attempt to model them explicitly. Instead, we initiate the BMU manually by perturbing the following fixed point of (\ref{PDE})
\begin{equation}  \left\{ 
              \begin{array}{llll}\label{BC}
               u_1(\boldsymbol{x},t)&=&u_{1,ss} \\
               u_2(\boldsymbol{x},t)&=&u_{2,ss}  \\
               \phi_R(\boldsymbol{x},t)&=&0  \\
               \phi_O(\boldsymbol{x},t)&=&0  \\
               z(\boldsymbol{x},t)&=&100.            \end{array}
       \right.
\end{equation} 
To initiate the BMU we proceed now as follows. We leave the osteoclast field at steady-state $u_{1,ss}$ everywhere except for a confined region $U$ where we add a few active cells $u_{1,pert}(\boldsymbol{x})>0$ for $\boldsymbol{x}\in U$. We assume that there are initially no active osteoblasts and that their density equals $u_{2,ss}$ everywhere. This is consistent with the assumption of the bone marrow being a precursor reservoir. The initial RANKL field is of great importance for the model because it is responsible for both targeted and strain-derived steering of the BMU. In fact, since neither the strain fields nor the osteocytes (which are responsible for the mechanotransduction by means of RANKL expression) are modeled explicitly as state-variables, possible damage sites and the principal stress directions have to be included in form of local perturbations of $\phi_{R,pert}(\boldsymbol{x})$. Finally, we assume that there is no OPG present in the initial system and that the bone density is at 100\% . In summary, the \emph{initial conditions} are given by:
\begin{equation} \left\{ 
              \begin{array}{lll}\label{IC}
                u_1(\boldsymbol{x},t=0)&=&u_{1,ss}+u_{1,pert}(\boldsymbol{x})  \\
               u_2(\boldsymbol{x},t=0)&=&u_{2,ss}  \\
                \phi_R(\boldsymbol{x},t=0)&=&\phi_{R,pert} (\boldsymbol{x}) \\
                \phi_O(\boldsymbol{x},t=0)&=&0 \\
                z(\boldsymbol{x},t=0)&=&100                \end{array}
       \right. \qquad \boldsymbol{x}\in \Omega.
\end{equation}
Since bone remodeling is a local process, we choose the domain large enough to avoid interactions of the BMU with the boundary. Note that for the BMU life spans considered hereinafter, \emph{large enough} means at least one order of magnitude longer than the cutting cone. The corresponding Dirichlet \emph{boundary conditions} for (\ref{PDE}) are given in (\ref{BC}) with $\boldsymbol{x}\in \partial \Omega$. \\
\indent Three comments regarding equations (\ref{PDE}) - (\ref{IC}) are in order. First, we draw attention to the fact that the osteoblast and the bone density equations are ordinary differential equations and can be integrated explicitly. In particular, for the $u_2$ equation we get
 \begin{equation}\label{OBtemp}
 u_{2}(\boldsymbol{x},t)=u_{2,ss}e^{-\beta_2 t}+\alpha_2 \int_{0}^{t}e^{\beta_2 (s-t)}u_1^{g_{12}}(\boldsymbol{x},s) ds.
 \end{equation}
 Second, the Heaviside function introduces a discontinuity into the equations, raising questions about the well-posedness of the PDE.  It can be seen that the point $(u_{1,ss}, u_{2,ss}, 0,0)$ is not a stable fixed point of the system. In the situations of interest, however, $y_1$ cannot be zero unless  $\phi_R$ is as well, since the active osteoclasts are only present in the cutting cone. Hence, we do not encounter issues of non-uniqueness. The questions of uniqueness and stability of the PDE system for the general situation are of interest, and are the subject of current work.
 
Third, we expect the osteoclast field $u_1$ and the RANKL field $\phi_R$ to inherit the singular behavior of the Heaviside function in (\ref{PDE}). In addition, the RANKL field also suffers from porous diffusion effects, which themselves are known to exhibit singular behavior. If the initial RANKL field is compactly supported in a region with a smooth boundary, this free surface may develop local corners and cusps in the course of the simulation \cite{Ebmeyer2005Regularity}. Indeed, if we allow $\phi_R$ to become negative (dropping below some threshold), very little can be said about the regularity of the ensuing PDE. This is an interesting question in its own right and will affect how computations may be performed. However, at this present juncture, we restrict ourselves to non-negative RANKL fields.

\section{The 1D model}\label{1Dmodel}

Due to the complexity of the model and the multitude of unknown parameters, we look at the 1D version of  (\ref{PDE}) - (\ref{IC}) before proceeding to the computationally more expensive 2D case. Note that in one dimension ($n=1$), the differential operators simplify as $\nabla\mapsto \partial_x$ and $\Delta\mapsto \partial_{xx}$. Before solving the system numerically, we first use some ideas of scaling theory to get a better understanding of physiological and pathological remodeling regimes as well as the corresponding parameter sets. 

\subsection{Parameter estimation and sensitivity analysis}\label{paramsens}
The primary goal after having established the model (\ref{PDE}) - (\ref{IC}) is to identify a \--- not necessarily unique \--- set of parameters that corresponds to a physiological remodeling regime. Once this is achieved, various combinations of parameters can then be modified to study the emergence of pathologies. Ideally, the physiological parameter set could be estimated on the basis of experimental data. However, since almost none of our 23 parameters can be matched with experimental findings, we are forced to adopt a different strategy. First, we consider the purely temporal model (\ref{temp1alpha}) and follow the reasoning in \cite{Komarova2003Mathematical} to obtain meaningful values. In particular, the values for $\beta_i$ can be estimated from experimental findings about the corresponding life spans of bone cells. Also, it is shown that the value of $g_{12}$ leads to unstable results outside of the interval $[0.1, 4]$ and that $g_{11}$ determines the overall dynamics of the cell populations. These facts, together with an estimation of the time delays ($t_R$, $t_O$) \cite{Baldock2006Vitamin, Thomas2001Changing, Gori2000Expression} and the aim of having a ratio of $u_{2,ss}/u_{1,ss}\approx100$ \cite{Parfitt1994Osteonal}, lead to the choice of $\alpha_i$, $\beta_i$,  $g_{ij}$, $t_R$ and $t_O$  found in (\ref{known1D}). The remaining parameters cannot be matched with experimental data and we determine their physiological values \emph{a posteriori}. More precisely, we fix the parameters in (\ref{known1D}), run simulations (as described in Section \ref{1Dexp}) and vary the remaining unknown parameters until the following two criteria are matched: first, the numerical solution has to coincide spatially and temporally with the global dynamics of \textit{in vivo} observations and second, the cutting cone has to stay compact  and move at a fairly constant speed. The outcome of this approach leads to the values summarized in  (\ref{unknown1D}).\\
\indent Now that the physiological set is determined, we can investigate the sensitivity of the model to parameter changes. To alleviate this task, we decide to focus on pathologies in the RANK/RANKL/ OPG pathway only. In other words, we consider the (\ref{known1D}) parameters from now on as \emph{fixed parameters}Ê and merely consider variations in the remaining \emph{free parameters} of (\ref{unknown1D}). However, a systematic sensitivity analysis of the 13 free parameters is still a rather unrealistic undertaking. Instead, we employ a scaling approach to analyze which parameters are able to destabilize the physiological regime and lead to the emergence of pathologies. The essence of scaling theory is to non-dimensionalize the equations by finding well-chosen scales for all the state-variables as well as the time and space variables. This leads to scaled equations where each term decomposes into the product of a dimensional coefficient representing the term's magnitude and a dimensionless factor of order of unity. Once this is achieved, it is possible to rewrite the equation in a dimensionless form where all the non-dimensional factors  are now preceded by so-called dimensionless groups that contain all the information about the terms' magnitudes. The dimensionless 1D version of (\ref{PDE}) reads
\begin{equation}  \left\{ 
              \begin{array}{lll}\label{ndPDE}
\partial_{\tilde{t}} \tilde{u}_1&=&G_1 \tilde{u}_{1}^{g_{11}} -G_2 \tilde{u}_1 - G_{3a} (\tilde{y}_1\partial_{\tilde{x}\tilde{x}} \tilde{\phi}_R) - G_{3b} (\partial_{\tilde{x}} \tilde{u}_1 \partial_{\tilde{x}} \tilde{\phi}_R)+G_4\frac{\tilde{ \phi}_R}{\tilde{\lambda}+\tilde{\phi}_R} \  \theta(\tilde{y}_1)\tilde{ u}_1 \\
\partial_{\tilde{t}} \tilde{u}_2&=&G_5 \tilde{u}_{1}^{g_{12}} -G_6 \tilde{u}_2 \\
\partial_{\tilde{t}} \tilde{\phi}_R &=& G_7 \,\tilde{ y}_{2,\tilde{t}_R} + G_8 \partial_{\tilde{x}\tilde{x}} (\tilde{\phi}_{R}^{\epsilon_R})- G_9 \frac{ \tilde{\phi}_R}{\tilde{\lambda}+\tilde{\phi_R}} \  \theta(\tilde{y}_1)\ \tilde{u}_1- G_{10}\tilde{\phi}_R\tilde{\phi}_O \\
\partial_{\tilde{t}} \tilde{\phi}_O&=&G_{11} \, \tilde{y}_{2,\tilde{t}_O}   +G_{12} \partial_{\tilde{x}\tilde{x}}(\tilde{\phi}_O^{\epsilon_O})- G_{13} \tilde{\phi}_R\tilde{\phi}_O\\
\partial_{\tilde{t}} \tilde{z}&=&-G_{14}\tilde{ y}_1+G_{15}\ \tilde{y}_2. 
                                 \end{array}
       \right. 
\end{equation}
The dimensionless groups $G_i$ and the corresponding scales can be found in Appendix \ref{dimgroups}. Note that all the state variables $\tilde{u}_i$, $\tilde{\phi}_\omega$, $\tilde{z}$ as well as $\tilde{x}$ and $\tilde{t}$ are now dimensionless and we can directly compare the various terms to determine their relative importance. In other words, we are now able to look for the dimensionless groups and parameters whose perturbations have a big impact on the model's regime. \\
\indent From a biological point of view, the most significant quantity is the bone mass density $\tilde{z}(\tilde{x},\tilde{t})$. It contains the key information about the outcome of the remodeling process, i.e. it determines whether we have excessive, normal or insufficient remodeling of the bone tissue. Since the outcome of the bone mass balance is determined by the activities of osteoclasts and osteoblasts respectively, we have to focus primarily on the dynamics of $\tilde{u}_1$ and $\tilde{u}_2$. However, bearing in mind that $\tilde{u}_2$ only depends on $\tilde{u}_1$ and that the \textit{fixed} parameters are kept at physiological values, we are assured that the osteoclasts will recruit enough osteoblasts to replace the resorbed tissue. In other words, the key players in the remodeling process are the osteoclasts and at this point, we do not have to worry about the osteoblasts. The only restriction to bear in mind is that the number of cells admissible per area is limited due to the cells' finite sizes; we ensure this by only considering parameter ranges that respect the spatial limitation. Osteoclasts are governed by the competition of $G_3$ (magnitude of migration) and $G_4\frac{\Phi_R}{\lambda+\Phi_R}$ (magnitude of stimulation by RANKL) and we define their ratio as (refer to Appendix \ref{dimgroups} for the scales)
\begin{align}\label{ratio1}
\Gamma_1:=\frac{G_3}{G_4}\left(1+\frac{\lambda}{\Phi_R}\right)\approx \frac{\zeta Y_1 \Delta\Phi_R}{k_1 U_1 L_1^2 }\left(1+\frac{\lambda}{\Phi_R}\right)=\frac{\zeta Y_1 }{k_1  U_1 L_1^2 }\, min\left\{\Phi_R, L_1\sqrt{\frac{k_2 U_1}{\zeta(1+\frac{\lambda}{\Phi_R})}}\right\}\left(1+\frac{\lambda}{\Phi_R}\right).
\end{align}
Physiological remodeling only occurs if the two terms are well-balanced, $\Gamma_1\approx 1$. A first pathological scenario corresponds to $\Gamma_1\gg 1$, i.e. the BMU moves much faster than it can nourish its population and dies out. On the other hand, if $\Gamma_1\ll1$, we have too many osteoclasts produced in the cutting cone and hence too many osteoblasts recruited in the back of the BMU. Depending on the RANKL and OPG production rates, this can lead to an excessive production of RANKL which in turn creates more osteoclasts etc. 
This \textit{positive feedback loop} in the closing zone can be investigated by means of the $\phi_R$ equation. A poor balance of RANKL production and its inhibition by OPG can lead to the described dysfunction in the closing cone zone of the BMU. More precisely, we are interested in the ratio of the production of RANKL by osteoblasts ($G_7$) and its inhibition by OPG binding ($G_{10}$):
\begin{align}\label{ratio2}
 \Gamma_2:=\frac{G_7}{G_{10}}\approx \frac{a_R Y_2}{k_3 \tilde{\Phi}_R \Phi_O }=\frac{a_R \beta_2}{ a_O k_3 \tilde{\Phi}_R }.
\end{align}
A high ratio $\Gamma_2\gg 1$ leads eventually to a singular behavior of the model (blow-up of the cell populations). Yet another pathological mechanism involves the OPG field in the closing zone and can lead to an early termination of the BMU. More precisely, if we have high production of OPG ($G_{11}$) in combination with low RANKL inhibition ($G_{13}$), i.e. if 
\begin{displaymath}
\Gamma_3=\frac{G_{11}}{G_{13}}\approx \frac{a_O Y_2}{k_3 \Phi_O \tilde{\Phi}_R }=\frac{\beta_2 }{k_3 \tilde{\Phi}_R }
\end{displaymath}
is very big, $\Gamma_3\gg1$, then the OPG field can possibly outrun the cutting cone and inhibit the RANKL field ahead of the BMU. The resulting lack of stimulation for the osteoclasts of the cutting cone can then lead to the extinction of the BMU. Obviously, this phenomenon only occurs if the diffusion is high relatively to the BMU speed.

\subsection{Numerical Experiments in 1D}\label{1Dexp}
\begin{figure}[!h]
\vspace{-4cm}
\begin{center}
  \includegraphics[width=0.75\textwidth]{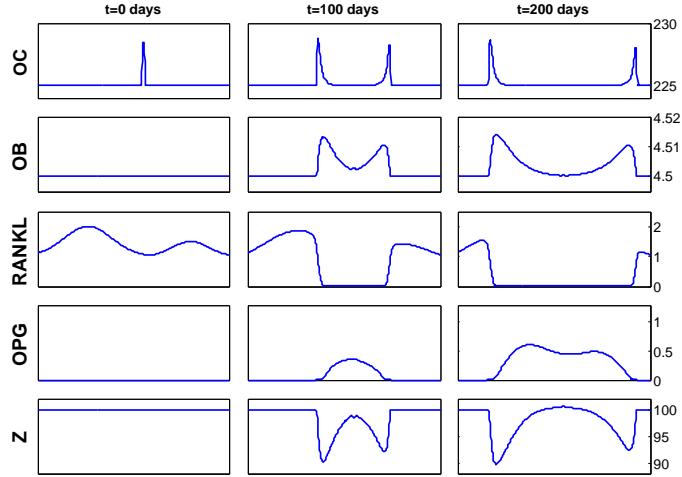}  
	\vspace{-3.7cm} \caption{\label{fig: 1Dpaperpic4}\footnotesize \textbf{Physiological remodeling II.}  OC=osteoclasts, OB=osteoblasts, Z=bone mass. The length of the domain is 15 mm and the OB scale is to be multiplied by $10^4$. Note that the remodeling mechanism is adaptive: the higher RANKL peak at $t=0$ leads to more remodeling, see Z at $t=200$. Parameter set and corresponding $\Gamma_i$ as in Figure \ref{fig:1Dpaperpic1} .}
\end{center}
\end{figure}

Note that the cutting cone of resorbing osteoclasts stays well-confined during the whole remodeling process and the BMU remodels a length of approximately 5 mm in 6.5 months. Therefore, the simulation satisfies our criteria for a physiological regime and validates the choice of parameters. Calculating the ratios defined in Section \ref{paramsens}, we get $\Gamma_1=\,0.83$, $\Gamma_2=1.1\cdot 10^{-3}$ and $\Gamma_3=2.7\cdot 10^{-3}$.
This is consistent with the previous discussion of parameter sensitivity. Indeed, $\Gamma_1\approx 1$ corresponds to a well regulated resorption activity, $\Gamma_2\ll1$ indicates a well-balanced RANKL distribution in the closing zone which is necessary for a confined cutting cone, and $\Gamma_3\ll1$ confirms that there is no risk of early termination due to excessive OPG production and diffusion. Finally, we point out that the scale estimations in Appendix A are in agreement with the simulation in Figure \ref{fig:1Dpaperpic1}.\\
\indent Using the same set of physiological parameters, we investigate now the situation where a BMU starts off in the middle of two zones of high RANKL concentration (this corresponds e.g. to the situation of two adjacent micro-fractures). Figure \ref{fig: 1Dpaperpic4} illustrates how the cutting cone splits into two parts and remodels each zone separately. 
In particular, the BMU remodeling the higher peak is more active as can be seen in the bone density evolution. In other words, the remodeling is adaptive: the bigger the damage and hence the RANKL expression, the higher the turnover in bone tissue. \\
\indent  The remainder of this section is dedicated to pathologies.  A first type of BMU malfunctioning is excessive bone remodeling and can be induced by two different imbalances. 
\begin{figure}[!h]
\vspace{-4cm}
\begin{center}
  \includegraphics[width=.75\textwidth]{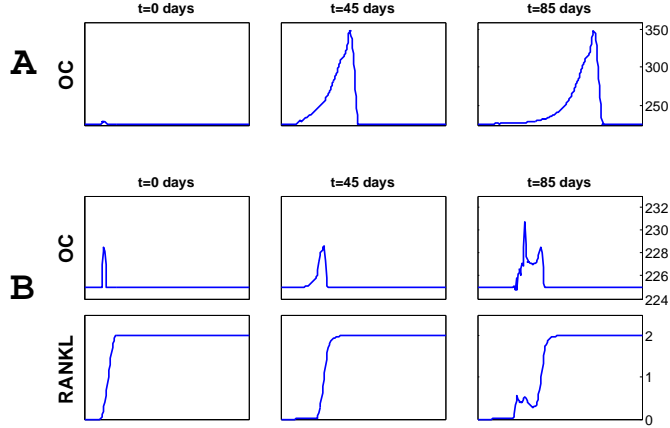}
	\vspace{-3.7cm} \caption{\label{fig:1Dpaperpic2}\footnotesize \textbf{Excessive remodeling.}  OC=osteoclasts; length of the domain is 10 mm. \textbf{A} Increased osteoclast recruitment and lower RANK-RANKL binding saturation lead to a larger but compact cutting cone in a stable regime. The ratios are $\Gamma_1=2.6\cdot 10^{-2}$, $\Gamma_2=1.3\cdot 10^{-3}$ and $\Gamma_3=3.2\cdot 10^{-3}$. The parameter set is given in (\ref{known1D}) and (\ref{unknown1D}) except for $\lambda=2$ and $k_1=9\cdot 10^{-2}$. \textbf{B} Very low OPG production by osteoblasts in the closing zone lead to a slow and unconfined cutting cone. Positive feedback leads to instability in the closing zone. The ratios are $\Gamma_1=52.2$, $\Gamma_2=1.3\cdot 10^{-3}$ and $\Gamma_3=1.7\cdot 10^{-3}$. The parameter set is given in (\ref{known1D}) and (\ref{unknown1D}) except for $a_O=2\cdot 10^{-8}$ after $t=60$ days ($a_O$ is kept high in the beginning to avoid numerical instabilities in the initiation zone).}
\end{center}
\end{figure}
If we decrease the ratio of osteoclast migration versus stimulation, i.e. if we choose the \emph{free parameters} such that $\Gamma_1\ll1$, then more osteoclasts and hence osteoblasts are recruited and therefore the amount of old bone tissue that gets remodeled is expected to be much higher. If we simultaneously ensure that the feedback loop parameter is small, $\Gamma_2\ll1$, we can avoid instabilities in the closing zone and expect an overall stable regime. These predictions are confirmed in the experiment illustrated in Figure \ref{fig:1Dpaperpic2}\emph{A}. Note in particular that the cutting cone, even though much longer, stays confined and no instabilities occur. However, instabilities can no longer be avoided if excessive remodeling is caused by unbalanced RANKL/OPG production in the closing zone. In order to illustrate this, we pick a parameter set such that $\Gamma_1\approx1$  but $\Gamma_2\gg1$. As shown in Figure \ref{fig:1Dpaperpic2}\emph{B}, the cutting cone is normal, but the excessive RANKL production in the closing zone leads to recruitment of a new generation of osteoclasts \emph{behind} the cutting cone. These osteoclasts attract in turn more osteoblasts which produce more RANKL, and the resulting positive feedback loop leads to well-visible instabilities. \\
\indent Yet another pathological scenario is the early termination of the remodeling process, i.e. the extinction of the BMU before its mission is accomplished. Here too, we distinguish two different causes.
\begin{figure}[!]
\vspace{-4cm}
\begin{center}
  \includegraphics[width=0.75\textwidth]{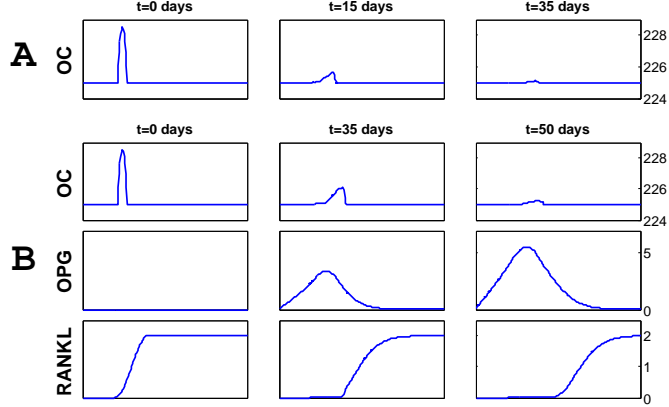}
\vspace{-3.7cm}	\caption{\label{fig:1Dpaperpic3}\footnotesize \textbf{Insufficient remodeling.}  OC=osteoclasts; length of the domain is 5 mm. \textbf{A} Decreased osteoclast recruitment and higher RANK-RANKL binding saturation lead to a vanishing cutting cone. The ratios are $\Gamma_1=10.5$, $\Gamma_2=7.2\cdot 10^{-4}$ and $\Gamma_3=1.8\cdot 10^{-3}$. The parameter set is given in (\ref{known1D}) and (\ref{unknown1D}) except for $\lambda=20$ and $k_1=3\cdot 10^{-4}$. \textbf{B} High production and diffusion of OPG leads to annihilation of the RANKL ahead of the BMU and lack of stimulation leads to BMU extinction. The ratios are $\Gamma_1=0.58$, $\Gamma_2=1.1\cdot 10^{-4}$ and $\Gamma_3=10.5$. The parameter set is given in (\ref{known1D}) and (\ref{unknown1D}) except for $a_O=1\cdot 10^{-2}$, $k_3=1.5\cdot 10^{-2}$ and $a_R=10^{-7}$.}
\end{center}
\end{figure}
If we choose $\Gamma_1\gg1$, then according to our discussion in Section \ref{paramsens} the osteoclast population will die out due to deficient stimulation. Consequently, the whole BMU slowly disappears, see Figure \ref{fig:1Dpaperpic3}\emph{A}. But early termination is also possible if osteoclasts respond well to RANKL stimulation: if the OPG production by osteoblasts largely exceeds the RANKL expression ($\Gamma_3\gg1$) and if the OPG diffusion is very high, then the excess of fast spreading OPG reaches the RANKL ahead of the cutting cone and annihilates the osteoclast stimulation. Figure \ref{fig:1Dpaperpic3}\emph{B} illustrates how the resulting lack in BMU stimulation can lead to early termination of the remodeling process.

\section{The 2D Model}\label{2Dmodel}

We extend the model now to two space dimensions to gain a better insight into the dynamics of trabecular remodeling. Let $\Omega\in \mathbb{R}^2$ denote a rectangular domain representing the surface of a  flat trabecula.  The local cell densities of osteoclasts $u_1$ and osteoblasts $u_2$ are denoted by $u_i(\boldsymbol{x},t)$, where $\boldsymbol{x}=(x,y)\in\Omega$, $i=1,2$. The RANKL field is denoted by $\phi_R(\boldsymbol{x},t)$ and the OPG field by $\phi_O(\boldsymbol{x},t)$. The governing equations are still given by (\ref{PDE}) - (\ref{IC}) and $\nabla$ and $\Delta$ are now the 
Divergence and Laplace operators in 2D. Since the width of a trabecula is small in comparison to its length \cite{Muller1995Three}, and since the bone tissue is separated from the bone marrow through a canopy of bone lining cells \cite{Hauge2001Cancellous}, vertical losses of RANKL and OPG are negligible (see also section \ref{mathmod}). This then justifies the use of a two dimensional diffusion equation to model the spread of chemicals across the surface of the trabecula. Note that we use the nonlinear, porous version of diffusion because the trabecular surface is very irregular and diffusing chemicals constantly interact with the components of the bone matrix as well as adjacent bone cells. In the remainder of this section, we present two numerical experiments on trabecular remodeling in a physiological regime. The first experiment is a regular micro-fracture remodeling and the second one illustrates that OPG plays an important role of counterbalancing the effects of RANKL and hence as a regulator for BMU-internal cell dynamics. More 2D-experiments in both physiological and pathological regimes together with a more detailed biological analysis of the results can be found in the accompanying article \cite{Ryser2008Mathematical}. Note finally that even though the scaling approach adopted for the 1D case in Section \ref{paramsens} looses its general validity in 2D, it can still be used to narrow down the plausible parameter ranges.

\subsection{Numerical Experiments in 2D}
\begin{figure}[htbp] 
   \centering
   \includegraphics[width=.5\textwidth]{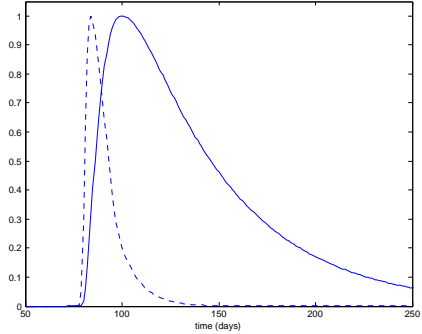} 
   \vspace{-2cm}
 \caption{Normalized population dynamics of active osteoclasts (dashed line) and active osteoblasts (solid line) at $x=3.7 mm$ in the experiment \emph{Physiological Remodeling I} (see Fig. \ref{fig:1Dpaperpic1}). We see that the passage time of the cutting cone is very short relatively to the time scale of the osteoblast dynamics. Therefore, one can assume that all the active osteoblasts are of  approximatively the same age.}
  \label{fig:OC_OB}
 \end{figure}

\begin{figure}[htbp] 
   \centering
   \includegraphics[width=.4\textwidth]{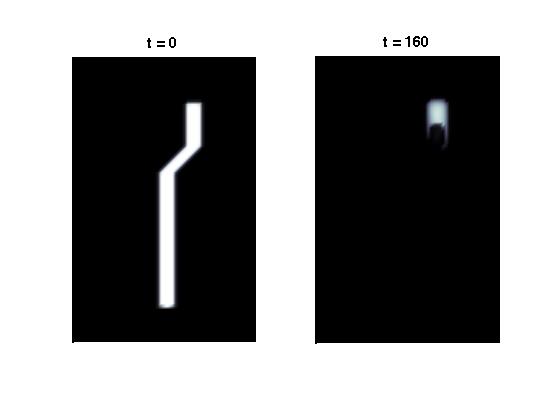} 
   \caption{\textbf{RANKL field simulating microfracture in trabecular bone}. Prior to the simulation, damaged osteocytes along the fracture express membrane-bound RANKL leading to the initial field at t=0 days (white indicates high concentration). At time t=160 days, the RANKL field has almost entirely disappeared after having bound to both OPG and RANK receptors on osteoclasts. Since RANKL is membrane-bound, the diffusion is very low. Length of domain is 3 mm.}  \label{fig:PP11}
 \end{figure}

\begin{figure}[!h] 
\vspace{-4cm}
 \begin{center}
    \includegraphics[ width=.8 \textwidth]{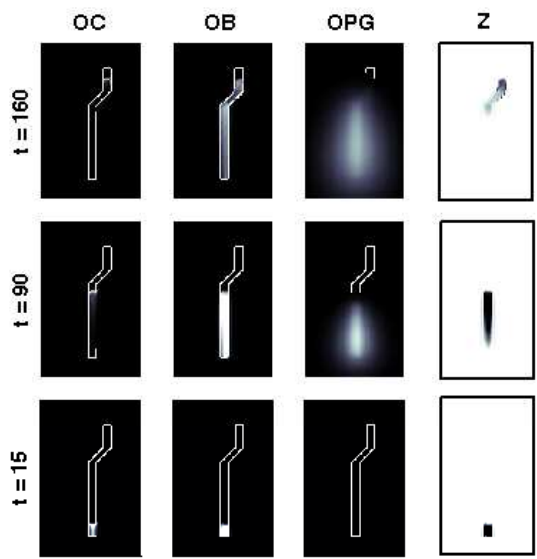}  
 \end{center} 
\vspace{-3.7cm}  \caption{\textbf{Steering of BMU along microfracture.} \emph{OC}: aggregation of osteoclasts (cutting cone) moving from the bottom of the domain to the top along the RANKL gradient.  \emph{OB}: osteoblasts, rebuilding the bone in the wake of the cutting cone. \emph{OPG}: diffusing OPG field. \emph{Z}: evolution of the bone mass density. Outline of initial RANKL field (micro-fracture) is highlighted for reference; length of domain is 3 mm; black corresponds to low, white to high concentrations.}   \label{fig:PP2}
\end{figure}

\begin{figure}[h!]
 \vspace{-4cm}
\begin{center}
  \includegraphics[width=.8\textwidth]{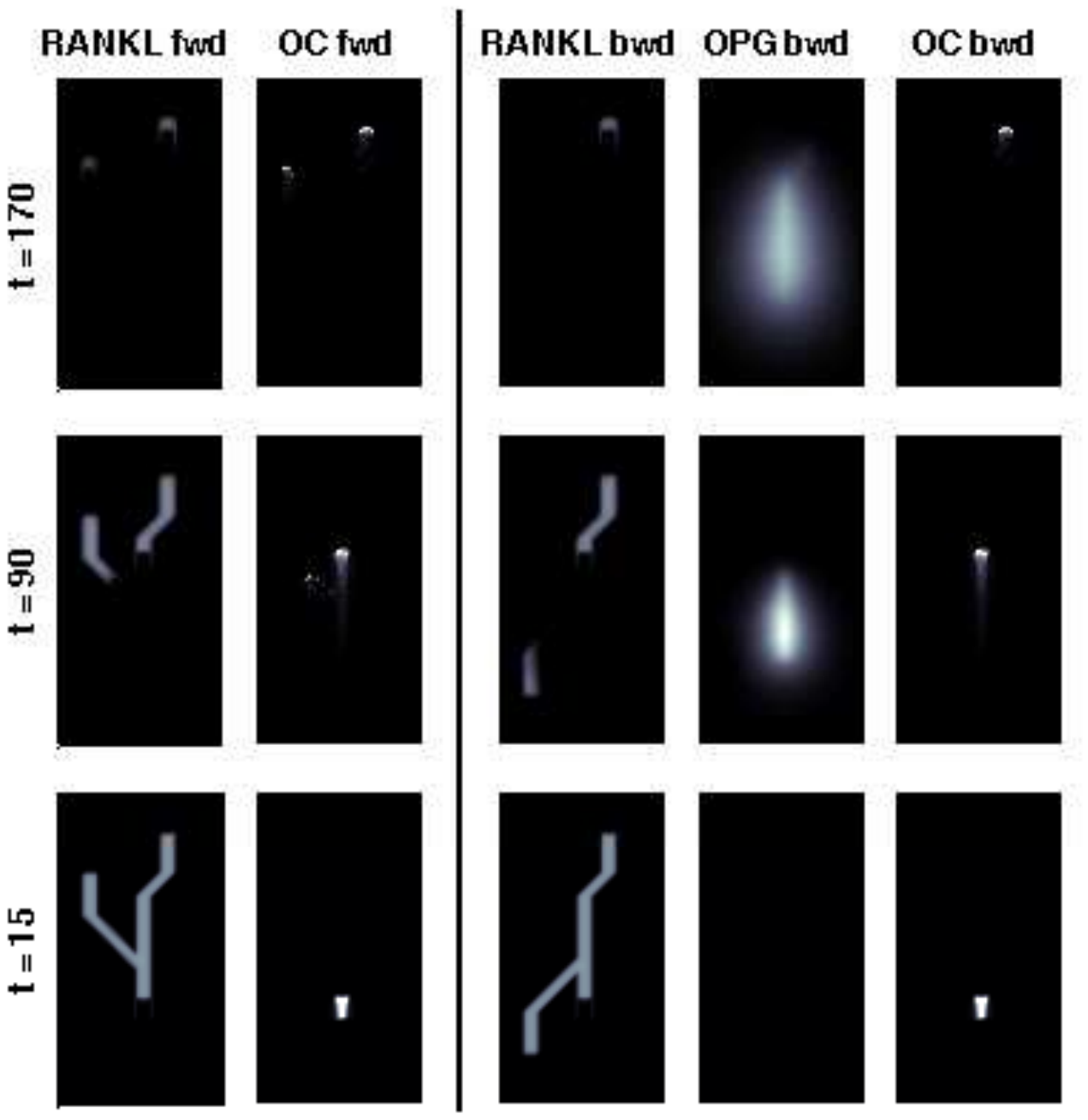}
  \end{center}
 \vspace{-3.6cm} \caption{\textbf{Forward versus backward branching}. In addition to the primary micro-fracture, secondary fractures branching at $45^{\circ}$ and $135^{\circ}$ are added (see \emph{RANKL fwd} and \emph{RANKL bwd}Ê at $t=15$). \emph{OC fwd}: Remodeling in the forward direction is successful, the BMU splits into two parts and remodels each branch separately. \emph{OC bwd}: Remodeling in the backward direction is unsuccessful, the BMU only remodels the primary branch. The reason for this is annihilation of RANKL by OPG along the secondary branch before the cutting cone reaches the branching area (see \emph{RANKL bwd}Ê and \emph{OPG bwd}). The resulting lack of osteoclast stimulation prevents the BMU from branching and the backward branch are not remodeled. Length of domain is 3 mm.}
   \label{fig:PP3}
\end{figure}

The following experiments are based on the model (\ref{PDE}) - (\ref{IC}), only the time delay terms in the RANKL and OPG equations are replaced by $a_\omega \, y_{2,t_{\omega}} \mapsto a_\omega  \, y_2(\boldsymbol{x},t)\, \Xi(\boldsymbol{x},t,t_\omega )$, where 
\begin{align}\label{delayalt}
 \Xi(\boldsymbol{x},t,t_\omega )=\left\{\begin{array}{l} 1 \qquad \text{ if } y_2(\boldsymbol{x},\delta)>0 \text{ for some } \delta\in[0,t-t_\omega ] \, \text{and}\, t > t_\omega\\ 0 \qquad \text{ otherwise. }\end{array} \right.
\end{align} 
This means that if at a certain location there exists an active osteoblast older than $t_{\omega}$, then all the active osteoblasts at the same  location produce the respective chemical, independent of their age. This particular source term is practically useful because it does not require the use of delay differential equation solvers and hence improves both the computational cost and the stability of the algorithm. Furthermore, it is a reasonable approximation to the original version of the delay term $y_{2,t_\omega}$ as shown in Figure \ref{fig:OC_OB}.  In fact, considering the passage time of the cutting cone in the case of a physiological 1D experiment shows that the latter is very short relatively to the time scale of the osteoblast dynamics. In other words, it is reasonable to assume that all the active osteoblasts at a specific location are of roughly the same age. In addition, the delay times $t_\omega$ are such that $e^{-\beta_2 t_\omega}\approx 1$, and we conclude that  $y_2(\boldsymbol{x},t)\, \Xi(\boldsymbol{x},t,t_\omega )$ is indeed a reasonable approximation for $ y_{2,t_{\omega}}$. 
 All the simulations are performed in \emph{Matlab} by means of a second order finite difference scheme in space and the built-in solver {\tt ode45} in time.\\
\indent First, we demonstrate the effect of RANKL on BMU steering along a micro-fracture. The mechanically damaged osteocytes adjacent to the fracture create a path of membrane-bound RANKL as depicted in Figure \ref{fig:PP11} at time $t=0$. 
In the course of the simulation, the RANKL-guided BMU remodels the fracture and the RANKL disappears due to RANK-RANKL binding, leading to the final snapshot after  $t=160$ days. We initiate the BMU by introducing a confined aggregation of active osteoclasts at the bottom of the microfracture at time $t=0$. The first panel in Figure \ref{fig:PP2} shows the subsequent motion of the cutting cone: the bright area represents the region of active osteoclasts which move towards the top of the fracture. 
The osteoblast dynamics are depicted in the second panel: osteoblasts are recruited by active osteoclasts and produce new bone matrix in the areas where the cutting cone has already resorbed the bone. The third panel shows the OPG field: it is produced by mature osteoblasts and hence lags the cutting cone. In particular, OPG is not membrane-bound and diffuses across the trabecula at a fairly high speed. The last panel depicts the evolution of the bone density. Note finally that the cutting cone stays well-confined and the BMU moves at constant speed over 2 mm in 5 months. This is in agreement with experimental observations \cite{Parfitt1994Osteonal} and thus provides a validation of the chosen physiological parameter set. \\
\indent The second experiment is an extension of the 1D experiment on the possibility of BMU branching in the case of multiple micro-fractures (see Figure \ref{fig: 1Dpaperpic4}) . More precisely, we want to find out if a BMU can split into two separate BMUs and if it can, then we want to investigate the existence of preferential branching directions. We start off with the initial RANKL field from the previous experiment and add a secondary branch which deviates by $45^{\circ}$ from the primary branch as shown in the snapshot \emph{RANKL fwd} at $t=15$ in Figure \ref{fig:PP3}.
 Again, an initial aggregation of osteoclasts is placed at the bottom of the microfracture and as time progresses, this cutting cone moves towards the top of the fracture. Similarly to the 1D experiment, the BMU splits into two individual parts which remodel both branches separately. Interestingly, if one repeats this experiment with the secondary branch deviating at $135^{\circ}$ rather than $45^{\circ}$ (see \emph{RANKL bwd}), the BMU remodels the primary but \emph{not} the secondary branch. In fact, the RANKL in the secondary branch is annihilated by OPG-RANKL binding before the cutting cone reaches the branching location, and in this way \--- due to the resulting lack of RANKL stimulation \--- the osteoclasts do not deviate from the primary branch (see \emph{OC bwd}). In summary, the OPG production in the back of the BMU prevents the BMU from turning around and remodeling the previously remodeled tissue. This is in good agreement with experimental results obtained by means of microcomputed tomography (MCT) imaging by \cite{Cooper2006Threedimensional}. For a more detailed discussion of this branching phenomenon we refer to \cite {Ryser2008Mathematical}.

\section{Conclusion and Outlook}\label{concl}

We have established a novel mathematical model of bone remodeling at cellular level. Based on a previous temporal model by Komarova et al. we developed step by step a spatio-temporal model describing both the osteoblast and osteoclast populations as well as the dynamics of the RANKL and OPG fields. The complete model has been shown to successfully recapitulate the overall dynamics of a single BMU as well as the distinct features of the cutting cone. Scaling was used to investigate the importance of the various model parameters and to motivate experiments on pathological remodeling.\\
\indent A strong feature of our model is the possibility to investigate the role of the spatial RANKL and OPG distributions in the osteoblast-derived paracrine control of osteoclasts. Even though there is a strong consensus in the experimental literature about the importance of the RANKL/OPG ratio, the following apparent inconsistency has to our knowledge not yet been addressed. In fact, the cutting and closing cones are spatially disconnected and hence osteoblasts appear when osteoclasts are already gone. So how can osteoblasts possibly play a key role in the osteoclast control$\,$? Our results show that the spatially distinct distributions of the RANKL and OPG fields provide the missing link: by expressing the diffusing RANKL-inhibitor OPG, osteoblasts have an indirect means of control over the activity of osteoclasts and hence the extent of remodeling and the direction of movement of the whole BMU.  
\indent The  2D version of the model is particularly suited to describe trabecular remodeling in the case where the local curvature of the trabeculae is negligible. Regarding cortical remodeling, it is likely that the 2D model provides a good qualitative approximation in the case when the BMU moves within the same plane of the cortical tissue. Nevertheless, in order to draw quantitative conclusions, a full 3D formulation together with a few modifications of the model assumptions become necessary.\\
\indent For future investigations, the model presented in this article provides a promising starting point. Besides an improvement of the numerical scheme and an extension to three dimensions for cortical remodeling, we also plan to improve our results by adding the natural decay rates as well as appropriate stochastic terms to the RANKL and OPG equations. In fact, since the production of messenger molecules by cells are subject to fluctuations, the use of noisy RANKL and OPG sources is expected to improve the model predictions in view of the often very irregular BMU evolutions observed \emph{in vivo}. Further model improvements might be achieved by describing  precursor cells as independent state variables and by including other important regulating factors such as Sclerostin, TGF-$\beta$ and PTH as state variables. However, the resulting increase in complexity would further compromise the balance between reliability and realism: the parameter-fitting for the current model already presents a substantial challenge and the addition of more unknown parameters would certainly not improve the model's quantitative reliability. Regarding the mechanical factors, model improvements can be achieved by taking into account the local curvature and by coupling the model to existing finite element models describing the elastic properties of the tissue.

\appendix

\section{Dimensionless groups and scale estimations}\label{dimgroups}
Due to the multiple time and length scales in the system (\ref{PDE}) as well as the occurrence of two different zones (the cutting zone and the closing zone), we have to abandon the idea of finding a consistent non-dimensional version of the original equation with a \emph{single} set of scales. However, we can still transform (\ref{PDE}) into a dimensionless equation  where the dimensionless factors preceded by the dimensionless groups are all of the order of unity  - all we have to do is to use different scales for different terms. Furthermore, the structure of the resulting equations  (\ref{ndPDE}) with respect to the time derivative implies that even if we cannot identify a single time scale for each equation separately, we can still compare the terms on the right hand side because the ratios of the form $G_{i}/G_{j}$ are independent of the time scale. The dimensionless groups in  (\ref{ndPDE}) are defined as
\begin{equation}\begin{array}{llll}\label{Gs}
G_1=T\, \left( \alpha_1 U_1^{g_{11}-1} \right) &
G_2= T \, \beta_2 &
G_{3a}= T\, \left( \frac{\zeta Y_1 \Delta\Phi_R}{U_1 L_{R}^2}\right)&
G_{3b}= T\, \left( \frac{\zeta Y_1 \Delta\Phi_R}{U_1 L_{R}L_{1}}\right)\\
G_4= T\, k_1 &
G_5= T\,\left( \frac{\alpha_2 U_1^{g_{12}}}{U_2}\right)&
G_6= T\, \beta_2 &
G_7=T\, \left(\frac{a_R Y_2}{\tilde{\Phi}_R}\right) \\
G_8=T\, \left(\frac{\kappa_R (\Delta \Phi_R)^{\epsilon_R}}{L_R^2 \tilde{\Phi}_R }\right) &
G_9=T\, \frac{k_2 U_1}{\tilde{\Phi}_R} &
G_{10}=T\, k_3 \Phi_O &
G_{11}=T\, \frac{a_O Y_2}{\Phi_O} \\
G_{12}=T\, \left(\frac{\kappa_O \Phi_O^{\epsilon_O-1}}{L_O^2}\right) &
G_{13}=T\, k_3 \tilde{\Phi}_R &
G_{14}=T\, \frac{Y_1 f_1}{Z} &
G_{15}=T\, \frac{Y_2 f_2}{Z} 

            \end{array}
\end{equation}
Except for $i\in\{8,12,15,16\}$, most of the terms $G_i$  play a significant role in only one of the two remodeling domains: $i\in\{1,2,3,4,5,6,9 \}$ in the cutting cone  and $i\in\{7,10,11,13\}$ in the closing cone. This has to be kept in mind when looking for the correct scales. The capital letters $U_i$, $Y_i$, $\Phi_R$, $\Phi_O$ and $Z$ are the scales of the corresponding state variables. $L_1$, $L_R$, $L_O$ are the length scales of the osteoclast, the RANKL and the OPG fields respectively. $\Phi_R$ and $\tilde{\Phi}_R$ scale the RANKL field at the tip of the cutting and in the back of the closing zone respectively. $\Delta\Phi_R$ is the difference of the RANKL concentration between the front and the back of the cutting zone. \\
\indent In the remainder of this section, the various scales are now briefly justified. Since we assume physiological remodeling conditions, the length of the cutting cone and the number of its constituent active cells are supposed to be preserved. Therefore, the length scale $L_1$ equals the initial length of the cutting cone and $Y_1\approx u_{1,pert}$ with $u_{1,pert}$ the initial perturbation added to the steady-state pool of passive cells at time $t=0$. Consequently, $U_1\approx u_{1,ss}+u_{1,pert}$. For the RANKL field we note first that the biggest change in concentration occurs in the cutting zone and hence the corresponding length scale is $L_R\approx L_1$. Since physiological remodeling excludes excessive RANKL production by osteoblasts, the scale is dictated by the initial conditions, $\Phi_R\approx max_{x\in\Omega} |\phi_R(x,t=0)|$. Next we estimate the passage time $T_p$ it takes the cutting cone to move across its own span:  $L_1$ is divided by the velocity of the osteoclasts to get $T_p=\frac{L_1^2}{\zeta \Delta\Phi_R}$. This expression allows us to eliminate $T_p$ in the estimation for $\Delta\Phi_R\approx T_p k_2 U_1 \frac{\Phi_R}{\lambda+\Phi_R}$ and we obtain, respecting the positivity requirement of the field, 
\begin{equation}\label{DeltaPhiR}
\Delta \Phi_R\approx\min\left\{\Phi_R,L_1 \sqrt{\frac{k_2 U_1}{\zeta (1+\frac{\lambda}{\Phi_R})}}\,\right\}.
\end{equation}
The remaining RANKL scale $\tilde{\Phi}_R$ is given by $\tilde{\Phi}_R\approx \Phi_R-\Delta \Phi_R + a_R (t_O-t_R) Y_2$. Using the time $T_p$ we get then for active osteoblasts $Y_2\approx T_p \alpha_2 Y_1=\frac{\alpha_2 L_1^2}{\zeta \Delta\Phi_R}Y_1$ and hence $U_2\approx u_{2,ss}+\frac{\alpha_2 L_1^2}{\zeta \Delta\Phi_R}Y_1$. Since the OPG field is generated by active osteoblasts with a life span of $1/\beta_2$, we get the estimates $\Phi_O\approx a_OY_2/\beta_2$ and $L_O\approx\frac{\zeta \Delta \Phi_R}{L_1 \beta_2}$. Finally, the bone mass is scaled with respect to $Z\approx 100$.

\section{Parameter Values for 1D Experiments}\label{paramval1D}
As discussed in Section \ref{paramsens}, we distinguish between \textit{fixed} and \textit{free} parameters. The former are unchanged throughout all the experiments and their numerical values are summarized in the following table.
\begin{equation}\label{known1D}
 \begin{array}{llll}
\alpha_1=9.49 \unit{day^{-1} mm^{-0.5}} & \alpha_2=4 \unit{day^{-1}}& \beta_1=0.2 \unit{day^{-1}}& \beta_2=0.02 \unit{ day^{-1}}\\
g_{11}=0.5 & g_{12}=1 & t_R=4 \unit{day} & t_O=8 \unit{day}
 \end{array}
\end{equation}
The \textit{free} parameters are changed from experiment to experiment. For the physiological experiments we use the following set.
\begin{equation}\label{unknown1D}
 \begin{array}{llll}
\zeta=7\cdot 10^{-4} \unit{mm^3\,mol^{-1}\, day^{-1}} & k_1=3\cdot 10^{-3} \unit{day^{-1}} & \lambda =5 \unit{mol\,mm^{-1}} \\ a_R=6\cdot10^{-5}\unit{mol\,day^{-1}}& 
a_O=1.5\cdot10^{-4}\unit{mol\,day^{-1}} & \kappa_R=3.16\cdot10^{-5}\unit{mm^{\epsilon_R+1}\,mol^{1-\epsilon_R}\,day^{-1}} \\ \epsilon_R=2.5 \qquad \epsilon_O=1 
& k_2=1\cdot10^{-3}\unit{mol\,day^{-1}} & k_3=1.2\unit{mm\,mol^{-1}\,day^{-1}}\\ 
\kappa_O=10^{-3}\unit{mm^{\epsilon_O+1}\,  mol^{1-\epsilon_O}\,day^{-1}} &   f_1=0.3\unit{g\,day^{-1}} & f_2=1.6\cdot 10^{-3}\unit{g\,day^{-1}}
 \end{array}
\end{equation}

\section{Parameter Values for 2D Experiments}\label{paramval2D}
\begin{equation}\label{param2D}
 \begin{array}{llll}
\alpha_1=30 \unit{day^{-1} mm^{-1}} & \alpha_2=4 \unit{day^{-1}}& \beta_1=0.1 \unit{day^{-1}}& \beta_2=0.02 \unit{ day^{-1}} \\
g_{11}=0.5 & g_{12}=1 & t_R=5 \unit{day} & t_O=15 \unit{day} \\
\zeta=10^{-5} \unit{mm^4\,mol^{-1}\, day^{-1}} & k_1=2.8\cdot 10^{-3} \unit{day^{-1}} & \lambda =50 \unit{mol\,mm^{-2}} & a_R=10^{-6}\unit{mol\,day^{-1}}\\ 
a_O=3\cdot10^{-4}\unit{mol\,day^{-1}} & \kappa_R=10^{-9}\unit{mm^{2\epsilon_R}\,mol^{1-\epsilon_R}\,day^{-1}} & \epsilon_R=3 \qquad \epsilon_O=1 
& k_2=4.6\cdot10^{-4}\unit{mol\,day^{-1}} \\ k_3=5\,10^{-3}\unit{mm^{2}\,mol^{-1}\,day^{-1}} &
\kappa_O=10^{-3}\unit{mm^{2\epsilon_O}\, mol^{1-\epsilon_O}\, day^{-1}} &   f_1=0.24\unit{g\,day^{-1}} & f_2=1.7\cdot 10^{-3}\unit{g\,day^{-1}}
 \end{array}
\end{equation}

\section*{Acknowledgements}
NN gratefully acknowledges the support of NSERC and the NSERC Accelerator Award program. SV gratefully acknowledges support from NSERC. The authors are indebted to Paul Tupper for helpful discussions and comments. We also thank the anonymous referees for their helpful suggestions which have lead to many improvements in the paper.

\bibliography{SIAM_1}
	\bibliographystyle{plain}

\end{document}